\documentclass[12pt]{article}
\usepackage{amssymb,amsmath,epsfig}

\begin{document}

\title{\bf Anisotropic Dark Energy and the Generalized Second Law of Thermodynamics}
\author{M. Sharif \thanks {msharif.math@pu.edu.pk} and Farida
Khanum \thanks {faridakhanum.math@yahoo.com}\\
Department of Mathematics, University of the Punjab,\\
Quaid-e-Azam Campus, Lahore-54590, Pakistan.}

\date{}

\maketitle
\begin{abstract}
We consider a Bianchi type $I$ model in which anisotropic dark
energy is interacting with dark matter and anisotropic radiation.
With this scenario, we investigate the validity of the generalized
second law of thermodynamics. It is concluded that the validity of
this law depends on different parameters like shear, skewness and
equation of state.
\end{abstract}

\section{Introduction}

The observational evidence that our universe is making a transition
from a decelerating phase to an accelerating is the major
development in cosmology. Supernova $Ia$ data \cite{1,2} gave the
first indication of the accelerated expansion of the universe. This
was confirmed by the observations of anisotropies in the cosmic
microwave background (CMB) radiations as seen in the data from
satellites such as Wilkinson Microwave Anisotropy Probe (WMAP)
\cite{3} and large scale structure \cite{4}. Astrophysical
observations indicate that accelerated expansion of the universe is
driven by an exotic energy with  large negative pressure which is
known as \emph{dark energy} (DE).

The data indicates that the universe is spatially flat and is
dominated by 76\% DE and 24\% other matter (20\% dark matter and 4\%
other cosmic matter). Despite all lines of observational evidence,
the nature of DE is still a challenging problem in theoretical
physics. Several models have been proposed such as, quintessence
\cite{5}, phantom field \cite{6}, tachyon field \cite{7}, quintom
\cite{8} and the interacting DE models, Chaplygin gas \cite{9},
holographic models \cite{10} and braneworld models \cite{11}, etc.
However, none of these models can be regarded as being entirely
convincing so far.

In black hole physics, the temperature and entropy are proportional
to the surface gravity at the horizon and the area of the horizon
\cite{12,13}, respectively. Also, the temperature, entropy and mass
of the black hole satisfy the first law of thermodynamics \cite{14}.
This leads to the relationship between the black hole thermodynamics
and the Einstein field equations. Jacobson \cite{15} derived the
field equations from the first law of thermodynamics for all local
Rindler causal horizons. Padmanavan \cite{16} formulated the first
law of thermodynamics on the horizon using the field equations for a
general static spherically symmetric spacetime. Later, this
relationship was developed in the cosmological context by taking
universe as a thermodynamical system bounded by the apparent
horizon. Cai and Kim \cite{17} showed the equivalence of the first
law of thermodynamics to the Friedmann equations with the
generalized second law of thermodynamics (GSLT) at the horizon.

Cosmological models in which DE interacts with dark matter and other
cosmic matter are well-known in the literature \cite{18}. The
discovery of black hole thermodynamics has led to the thermodynamics
of cosmological models. Bekenstein \cite{13} proved that there is a
relation between an event horizon and thermodynamics of black hole.
The event horizon of black hole is in fact the measure of entropy
which is generalized to the cosmological models so that each horizon
corresponds to an entropy. The GSLT is generalized in such a way
that the sum of the time derivative of each entropy must be
increasing. People studied the thermodynamics of different models in
which different cosmic constituents interact.

Wang et al. \cite{20} found that both the first and the second law
of thermodynamics break down at the event horizon. The first law of
thermodynamics is restricted to nearby states of local thermodynamic
equilibrium while the event horizon reflects global features of the
spacetime. Also, due to the existence of the cosmological event
horizon, the universe should be non-static in nature and, as a
result, the usual definition of the thermodynamical quantities on
the event horizon may not be as simple as in the static spacetime.
Finally, they have argued that as the event horizon is larger than
the apparent horizon so the universe bounded by the event horizon is
not a Bekenstein system.

There is a large body of literature \cite{21}-\cite{28e} dealing
with the thermodynamics of the universe bounded by the apparent
horizon and the validity of GSLT. Mazumder et al. \cite{29},
\cite{30} have explored the validity of GSLT using the first law of
thermodynamics and found some conditions on its validity. Debnath
\cite{31} investigated the validity of GSLT in the scenario of
Friedmann Robertson Walker (FRW) model in which DE interacts with
dark matter and radiation. Mubasher et al. \cite{32} proved the
validity of GSLT for all time, independent of geometry and equation
of state (EoS) parameter. In a recent paper \cite{32a}, the validity
of GSLT is explored in the Kaluza-Klein cosmology with modified
holographic DE.

Jacobs \cite{33} studied the spatially homogeneous and anisotropic
Bianchi type $I$ (BI) cosmological model with expansion and shear
but without rotation. He discussed anisotropy in the temperature of
CMB and cosmic expansion both with and without magnetic field.
Akarsu and Kilinc \cite{34} investigated anisotropic BI models in
the presence of perfect fluid and minimally interacting DE with
anisotropic EoS parameter. They found that anisotropy of the DE did
not always promote the anisotropy in the expansion.

In this paper, we take the BI model in which anisotropic DE
interacts with isotropic dark matter as well as anisotropic
radiation and investigate the validity of GSLT. The plan of this
paper is the following: in next section, we take the BI model as the
representation of the universe and evaluate shear parameters
$(R,S)$, skew parameters $(\delta,~\gamma,~\zeta,~\xi)$. Also, we
formulate the corresponding field equations, the generalized
Friedmann equation and equations of continuity in the effective
field theory. Section \textbf{3} presents the analysis of the
validity of GSLT. The last section concludes the paper results.

\section{Anisotropic DE Interacting with Two Fluids}

The metric representation of the BI universe is given as
\begin{equation}\label{1}
ds^{2}=-dt^{2}+A^2(t)dx^{2}+B^2(t)dy^{2}+C^2(t)dz^{2},
\end{equation}
where $A,~B,~C$ are scale factors along the $x,~y,~z$ axes
respectively. When $A=B=C$, the BI model reduces to the flat FRW
model. Thus BI is the generalization of the flat FRW model. The
energy-momentum tensor is defined as
\begin{equation}\label{2}
T^{\mu}_{\nu}=T^{\mu}_{(m)\nu}+T^{\mu}_{(\Lambda)\nu}+T^{\mu}_{(\chi)\nu},
\end{equation}
where the subscripts $m,~\Lambda,~\chi$ denote dark matter,
anisotropic DE and anisotropic radiation, respectively, and
corresponding energy momentum tensors are given as follows:
\begin{eqnarray}\label{3}
T^{\mu}_{(m)\nu}&=&(-\rho_{m},P_{m},P_{m},P_{m}),\\\label{4}
T^{\mu}_{(\Lambda)\nu}&=&(-\rho_{\Lambda},P_{(\Lambda)x},P_{(\Lambda)y},P_{(\Lambda)z}),\\\label{5}
T^{\mu}_{(\chi)\nu}&=&(-\rho_{\chi},P_{(\chi)x},P_{(\chi)y},P_{(\chi)z}).
\end{eqnarray}
The corresponding EoSs for these components are
\begin{eqnarray}\label{6}
P_{(\Lambda)x}&=&\rho_{\Lambda}\omega_{(\Lambda)x},\quad
P_{(\Lambda)y}=\rho_{\Lambda}\omega_{(\Lambda)y},\quad
P_{(\Lambda)z}=\rho_{\Lambda}\omega_{(\Lambda)z},\\\label{7}
P_{(m)x}&=&\rho_{m}\omega_{(m)x},\quad
P_{(m)y}=\rho_{m}\omega_{(m)y},\quad
P_{(m)z}=\rho_{m}\omega_{(m)z},\\\label{8}
P_{(\chi)x}&=&\rho_{\chi}\omega_{(\chi)x},\quad
P_{(\chi)y}=\rho_{\chi}\omega_{(\chi)y},\quad
P_{(\chi)z}=\rho_{\chi}\omega_{(\chi)z}.
\end{eqnarray}

To parameterize the anisotropy of pressure, we define skewness
parameters. These parameters actually measure the deviation in
pressure along the $y$ and $z$-axis from the $x$-axis. The skewness
parameters for anisotropic DE are defined as \cite{35}
\begin{equation}\label{9}
3\delta=\frac{P_{(\Lambda)x}-P_{(\Lambda)y}}{\rho_{\Lambda}},\quad
3\gamma=\frac{P_{(\Lambda)z}-P_{(\Lambda)x}}{\rho_{\Lambda}}.
\end{equation}
Using Eqs.(\ref{6}) and (\ref{9}), we can write Eq.(\ref{4}) as
follows:
\begin{equation}\label{10}
T^{\mu}_{(\Lambda)\nu}=[-\rho_{\Lambda},\omega_{\Lambda}\rho_{\Lambda},
(\omega_{\Lambda}+3\delta)\rho_{\Lambda},(\omega_{\Lambda}+3\gamma)\rho_{\Lambda}].
\end{equation}
The skewness parameters for anisotropic radiation are
\begin{equation}\label{11}
3\zeta=\frac{P_{(\chi)x}-P_{(\chi) y}}{\rho_{\chi}},\quad
3\xi=\frac{P_{(\chi)z}-P_{(\chi) x}}{\rho_{\chi}}.
\end{equation}
With the help of Eqs.(\ref{8}) and (\ref{11}), Eq.(\ref{5}) turns
out to be
\begin{eqnarray} \label{12}
T^{\mu}_{(\chi)\nu}=[-\rho_{\chi},\omega_{\chi}\rho_{\chi},
(\omega_{\chi}+3\zeta)\rho_\chi,(\omega_{\chi}+3\xi)\rho_{\chi}].
\end{eqnarray}
Making use of Eqs.(\ref{10}) and (\ref{12}) in Eq.(\ref{2}), we have
\begin{eqnarray}\nonumber
T^{\mu}_{\nu}&=&[-\rho_{m},\omega_{m}\rho_{m},\omega_{m}\rho_{m},\omega_{m}\rho_{m}]
+[-\rho_{\Lambda},\omega_{\Lambda}\rho_{\Lambda},(\omega_{\Lambda}
+3\delta)\rho_{\Lambda},(\omega_{\Lambda}+3\gamma)\rho_{\Lambda}]\\\label{2(a)}
&+&[-\rho_{\chi},\omega_{\chi}\rho_{\chi},(\omega_{\chi}
+3\zeta)\rho_{\chi},(\omega_{\chi}+3\xi)\rho_{\chi}].
\end{eqnarray}
The corresponding field equations take the following form:
\begin{eqnarray}\label{13}
\frac{\dot{A}\dot{B}}{AB}+\frac{\dot{B}\dot{C}}{BC}+\frac{\dot{C}\dot{A}}{CA}
&=&[\rho_{m}+\rho_{\Lambda}+\rho_{\chi}],\\\label{14}
\frac{\ddot{B}}{B}+\frac{\ddot{C}}{C}+\frac{\dot{B}\dot{C}}{BC}
&=&-[\omega_{m}\rho_{m}+\omega_{\Lambda}\rho_{\Lambda}+\omega_{\chi}\rho_{\chi}],\\\label{15}
\frac{\ddot{A}}{A}+\frac{\ddot{C}}{C}+\frac{\dot{A}\dot{C}}{AC}
&=&-[\omega_{m}\rho_{m}+(\omega_{\Lambda}+3\delta)\rho_{\Lambda}
+(\omega_{\chi}+3\zeta)\rho_{\chi}],\\\label{16}
\frac{\ddot{A}}{A}+\frac{\ddot{B}}{B}+\frac{\dot{A}\dot{B}}{AB}
&=&-[\omega_{m}\rho_{m}+(\omega_{\Lambda}+3\gamma)\rho_{\Lambda}+(\omega_{\chi}+3\xi)\rho_{\chi}].
\end{eqnarray}

The continuity equations, ${T^{\mu\nu}}_{;\nu}=0$, are
\begin{eqnarray}\label{17}
\dot{\rho_{\Lambda}}+H\rho_{\Lambda}[3(1+\omega_{\Lambda})
+\delta(3-2R+S)+\gamma(3-2S+R)]=-Q',\\\label{18}
\dot{\rho_{m}}+3H\rho_{m}(1+\omega_{m})=Q.
\end{eqnarray}
Here $Q$ and $Q'$ are introduced because two principal components of
the universe are mutually interacting, which lead to some loss in
other cosmic component. We take this component as radiation for
which the continuity equation becomes
\begin{equation}\label{19}
\dot{\rho_{\chi}}+H\rho_{\chi}[3(1+\omega_{\chi})+\zeta(3-2R+S)+\xi(3-2S+R)]=Q'-Q.
\end{equation}
Here, we define the mean expansion rate as an average Hubble rate
$H$ by
\begin{equation}\label{26}
H=\frac{1}{3}(\frac{\dot{A}}{A}+\frac{\dot{B}}{B}+\frac{\dot{C}}{C}).
\end{equation}
The difference of expansion rates as the Hubble normalized shear
parameters $R$ and $S$ are given as \cite{37}
\begin{equation}\label{27}
R=\frac{1}{H}(\frac{\dot{A}}{A}-\frac{\dot{B}}{B}),\quad
S=\frac{1}{H}(\frac{\dot{A}}{A}-\frac{\dot{C}}{C}).
\end{equation}
Shear parameters are also deviation measuring parameters which
parameterize anisotropy in scale factors by taking expansion along
$x$-axis as standard. Here, all expansion rates are positive and
$R>-3,~S<3$. This is also true either for $R=0$ or $S=0$. The
generalized Friedmann equation takes the form \cite{35}
\begin{equation}\label{28}
H^{2}=\frac{\rho_{m}+\rho_{\Lambda}+\rho_{\chi}}{3(1-\frac{1}{9}(R^{2}+S^{2}-RS))}.
\end{equation}
We take \cite{36}
$Q'=\Gamma_{\Lambda}\rho_{\Lambda},~Q=\Gamma_{m}\rho_{\Lambda}$.
The ratios for energy densities as
\begin{equation}\nonumber
r_{1}=\frac{\rho_{m}}{\rho_{\Lambda}},\quad
r_{2}=\frac{\rho_{\chi}}{\rho_{\Lambda}}.
\end{equation}
The corresponding EoS parameters in the effective field theory will
become
\begin{eqnarray}\label{20}
\omega^{eff}_{\Lambda}&=&\omega_{\Lambda}+\frac{\Gamma_{\Lambda}}{3H}
+\frac{\delta(3-2R+S)+\gamma(3-2S+R)}{3},\\\label{21}
\omega^{eff}_{m}&=&\omega_{m}-\frac{\Gamma_{m}}{3Hr_{1}},\\\label{22}
\omega^{eff}_{\chi}&=&\omega_{\chi}+\frac{\Gamma_{m}-\Gamma_{\Lambda}}
{3Hr_{2}}+\frac{\zeta(3-2R+S)+\xi(3-2S+R)}{3}.
\end{eqnarray}
Consequently, the continuity equations (\ref{17})-(\ref{19}) turn
out to be
\begin{eqnarray}\label{23}
\dot{\rho_{\Lambda}}+3H\rho_{\Lambda}(1+\omega^{eff}_{\Lambda})&=&0,\\\label{24}
\dot{\rho_{m}}+3H\rho_{m}(1+\omega^{eff}_{m})&=&0,\\\label{25}
\dot{\rho_{\chi}}+3H\rho_{\chi}(1+\omega^{eff}_{\chi})&=&0.
\end{eqnarray}

\section{Generalized Second Law Of Thermodynamics}

Now we investigate the validity of the GSLT in BI universe bounded
by apparent horizon with size $L$ which coincides with Hubble
horizon in the case of flat geometry, i.e., $L=\frac{1}{H}$. The
first law of thermodynamics gives
\begin{equation*}
TdS=PdV+dE,\quad dS=\frac{PdV+dE}{T},
\end{equation*}
where $T,~S,~E$ and $P$ are the temperature, entropy, internal
energy and pressure of the system respectively. The corresponding
entropies will become
\begin{equation}\label{29}
dS_{\Lambda}=\frac{P_{\Lambda}dV+dE_{\Lambda}}{T},\quad
dS_{m}=\frac{P_{m}dV +dE_{m}}{T},\quad
dS_{\chi}=\frac{P_{\chi}dV+dE_{\chi}}{T},
\end{equation}
where $P_{\Lambda},~P_{\chi},~P_{m},~E_{\Lambda},~E_{\chi}$ and
$E_{m}$ are the pressures and internal energies of anisotropic DE,
anisotropic radiation and dark matter, respectively. We assume that
the system is in equilibrium, which implies that all the components
of the system have the same temperature given by \cite{13}
\begin{equation}\label{30}
T=\frac{K}{2\pi },
\end{equation}
where $K$ is the surface gravity of the black hole. At equilibrium,
the horizon and other components of the system have same
temperature. Thermodynamical quantities are related to the
cosmological quantities by the following relations:
\begin{eqnarray}\label{31}
P_{\Lambda}&=&\omega^{eff}_{\Lambda}\rho_{\Lambda},\quad
P_{\chi}=\omega^{eff}_{\chi}\rho_{\chi},\quad
P_{m}=\omega^{eff}_{m}\rho_{m},\\\label{32}
E_{\Lambda}&=&ABC\rho_{\Lambda},\quad E_{\chi}=ABC \rho_{\chi},\quad
E_{m}=ABC\rho_{m},
\end{eqnarray}
where $V=ABC$ is the volume of the system containing  all the
matter.

The entropy of the horizon is $S_{h}=\frac{kA^*}{4}$, where $A^*$ is
the surface area of the black hole and $k$ is the Boltzmann's
constant. Here $A^*={4\pi L^{2}}$ so that we have $S_{h}=k \pi
L^{2}$, which leads to
\begin{equation}\label{34}
\dot{S_{h}}=2k\pi L \dot{L}.
\end{equation}
Also, the time derivative of Eq.(\ref{29}) yields
\begin{equation}\label{35}
\dot{S_{\Lambda}}=\frac{P_{\Lambda}\dot{V}+\dot{E_{\Lambda}}}{T},\quad
\dot{S_{\chi}}=\frac{P_{\chi}\dot{V}+\dot{E_{\chi}}}{T},\quad
\dot{S_{m}}=\frac{P_{m}\dot{V}+\dot{E_{m}}}{T}.
\end{equation}
Using Eqs.(\ref{23})-(\ref{25}), (\ref{34}) and (\ref{35}), it
follows that
\begin{equation}\label{36}
\dot{S}_{total}=2k\pi L\dot{L},
\end{equation}
where $\dot{S}_{total}$ is the sum of the matter entropy and horizon
entropy. Now making use of Eqs.(\ref{23})-(\ref{25}) and (\ref{28}),
we obtain
\begin{equation}\label{37}
\dot{L}=-\frac{\dot{H}}{H^2}=\frac{L^{3}H\alpha}{2\beta}-\frac{L^{3}H^{2}[\dot{R}(2R-S)
+\dot{S}(2S-R)]}{18\beta},
\end{equation}
where
\begin{eqnarray}\label{38}
&&\alpha=(1+\omega_{\Lambda})\rho_{\Lambda}+(1+\omega_{m})\rho_{m}
+(1+\omega_{\chi})\rho_{\chi}\nonumber\\
&&+\frac{1}{3}[(\rho_{\Lambda}\delta+\rho_{\chi}\zeta)(3-2R+S)
+(\rho_{\Lambda}\gamma+\rho_{\chi}\xi)(3-2S+R)],
\end{eqnarray}
and
\begin{equation}\label{39}
\beta={1}-\frac{R^{2}+S^{2}-R S}{9}\geq0.
\end{equation}
Inserting this value of $\dot{L}$ in Eq.(\ref{36}), it follows
that
\begin{equation}\label{40}
\dot{S}_{total}=\frac{ k \pi H L^{4}\alpha}{\beta}-\frac{ k \pi
H^{2} L^{4} [\dot{R}(2R-S)+\dot{S}(2S-R)]}{9\beta}.
\end{equation}
This implies that if $\dot{S}_{total}\geqslant0$, then GSLT holds.
Thus the validity of GSLT depends upon the skewness, shear and
state parameters. From Eq.(\ref{40}), $\dot{S}_{total}\geqslant0$
implies that
\begin{equation*}
\frac{ k \pi H L^{4}\alpha}{\beta}-\frac{ k \pi H^{2} L^{4}
[\dot{R}(2R-S)+\dot{S}(2S-R)]}{9\beta}\geq0.
\end{equation*}
After substituting the values of $\alpha$ and $\beta$ in the above
equation, it follows that
\begin{eqnarray}\label{41}
&&9(\rho+p)\geq[\dot{R}(2R-S)+\dot{S}(2S-R)]-3[(\rho_{\Lambda}\delta+
\rho_{\chi}\zeta)(3-2R+S)\nonumber\\
&&+(\rho_{\Lambda}\gamma+\rho_{\chi}\xi)(3-2S+R)].
\end{eqnarray}
Now, we consider the following three interesting
cases involving the condition on different parameters:\\
\textbf{I.} $\beta=0$, \textbf{II.} $\beta>0$, \textbf{III.}
$R=S=\delta=\gamma=\xi=\zeta=0$.

\subsection*{Case I}

Here $H$ and $\dot{S}_{total}$ tend to infinity for $\beta=0$ from
Eqs.(\ref{28}) and (\ref{40}) respectively. It may happen for very
large time $(t\rightarrow\infty)$, i.e., when the expansion rate is
very high. In this case, all the useable energy in the universe will
be converted into another form of energy which is not useable. This
stage is also known as the heat death of the system. The heat death
of the universe is said to be a suggested fate of the universe. At
this time, all the thermodynamic free energy will be diminished from
the universe and motion or life cannot sustain any more. In the
language of physics, the entropy in the universe will reach to its
maximum value.

\subsection*{Case II}

In this case, by takeing $\beta>0$ in Eq.(\ref{40}), we observe that
the validity of this law depends on the skewness, shear and EoS
parameters. Further, if we remove anisotropy in the expansion, its
validity depends on skewness as well as equation of state
parameters, and if we remove anisotropy in fluid, then it depends on
shear and equation of state parameters. Hence, GSLT is conditionally
valid.

\subsection*{Case III}

When we take $R=S=\delta=\gamma=\xi=\zeta=0$, BI reduces to the flat
FRW model. Using these values in Eq.(\ref{41}), we have
$(\rho+p)\geq0$, which is the null energy condition in the FRW
universe. Consequently, Eq.(\ref{40}) leads to
$\dot{S}_{total}=k\pi(\rho+p)$ and hence $\dot{S}_{total}\geq0$ for
$(\rho+p)\geq0$. Thus we can say that GSLT holds for all time in
this case, if the null energy condition for the considered matter is
satisfied in the FRW universe. This case implies that the validity
of GSLT depends on the null energy condition which is not discussed
in \cite{32}. Therefore using results produced in this case, we can
say that the validity of GSLT depends on null energy condition and
background geometry.

\section{Conclusion}

This paper is devoted to study the validity of GSLT with the BI
universe model. We assume that anisotropic DE is interacting with
dark matter and anisotropic radiation. Using this scenario, we have
formulated the conditions under which GSLT is valid. We have
discussed three cases for its validity. For a particular value of
$\beta=0$, the first case provides the general validity of GSLT. In
this case, we see that at infinite expansion the heat death of
universe will take place. At this stage all types of motion and life
will be finished. The second case $\beta>0$ gives conditional
validity of this law. In the third case, BI reduces to the flat FRW
metric and the general validity of GSLT depends upon the condition
$(\rho+p)\geq0$, which is the null energy condition. We would like
to mention here that the validity of this law for the FRW universe
with the same scenario has been proved \cite{32} independent of
background geometry, EoS parameter and interacting scenario. This
analysis indicates that the validity of GLST for FRW depends on
geometry as well as null energy condition. We conclude that GSLT is
conditionally valid in the BI type universe with anisotropic perfect
fluid.

\end{document}